\documentclass[aps,prl,twocolumn,superscriptaddress,longbibliography]{revtex4-2}

\usepackage{graphicx}
\usepackage{bm}
\usepackage{hyperref}
\usepackage{amsmath}
\usepackage{amssymb}
\usepackage[table]{xcolor}
\usepackage{array}
\usepackage{multirow}
\usepackage[caption=false]{subfig}
\usepackage{physics}

\usepackage{amsthm}
\usepackage{bbm}
\usepackage{multirow}
\usepackage[colorinlistoftodos]{todonotes}
\usepackage[scr=boondoxo,frak=boondox,scrscaled=1.05]{mathalfa}
\usepackage{amsmath}

\def\bra#1{\langle#1|}
\def\ket#1{|#1\rangle}
\def\braket#1#2{\langle#1|#2\rangle}

\renewcommand\vec{\boldsymbol}

\newcommand\Asout{\bgroup\markoverwith{\textcolor{red}{\rule[0.5ex]{4pt}{0.6pt}}}\ULon}

\usepackage{ulem}

\begin{document}
\raggedbottom
\title{Compressible quantum liquid with vanishing Drude weight}
\author{Ahmed Abouelkomsan}
\affiliation{Department of Physics, Massachusetts Institute of Technology, Cambridge, Massachusetts 02139, USA}
\author{Nisarga Paul}
\affiliation{Department of Physics, Massachusetts Institute of Technology, Cambridge, Massachusetts 02139, USA}
\author{Ady Stern}
\affiliation{Department of Condensed Matter Physics, Weizmann Institute of Science, Rehovot 7610001, Israel}
\author{Liang Fu}
\affiliation{Department of Physics, Massachusetts Institute of Technology, Cambridge, Massachusetts 02139, USA}

\begin{abstract}
    We explore the possibility of quantum liquids that are {\it compressible} but   
    have \textit{vanishing} DC conductivity in the absence of disorder. 
    We show that the composite Fermi liquid emerging from strong interaction in a generic Chern band 
   has zero Drude weight, in stark contrast to normal Fermi liquids. Our work establishes the absence of Drude weight as the defining property of the composite Fermi liquid phase, which distinguishes it from the Fermi liquid or other types of non-Fermi liquids. Our findings point to a possibly wide class of gapless quantum phases with unexpected transport and optical properties.
\end{abstract}

\maketitle


Compressibility and conductivity are two fundamental properties of quantum states of matter. 
Compressibility is a thermodynamic property that describes the response of charge density to a change in the chemical potential. It is closely related to the charge gap: as long as the range of inter-particle interaction is not too long, gapless states are compressible while gapped states are incompressible. On the other hand, DC conductivity is a transport property that distinguishes conducting and insulating states. Although conceptually different, compressibility and conductivity often appear concurrently:  
band insulators are incompressible, while metals are compressible. 

In this work, we demonstrate the existence of compressible quantum liquids with electric transport properties that differ drastically from normal metals.
As a diagnostic for electrical conductivity in the absence of disorder, we turn to the Drude weight which characterizes the response of a given system under adiabatic flux insertion, physically amounting to applying a DC electric field \cite{laughlin_quantized_1981,kohn_theory_1964}. A finite Drude weight is expected when level crossings between the ground state and excited states appear under flux insertion, leading to energy absorption.
We present theoretical argument and numerical evidence for the absence of Drude weight in certain strongly correlated quantum liquids. These ``Drudeless metals'' are sharply distinguished from ordinary metals which become perfect conductors in the clean limit, as well as  
Anderson insulators  which are compressible but insulating because of disorder-induced localization.

Our study is motivated by the recent prediction of a non-Fermi liquid state in twisted bilayer semiconductor $t$MoTe$_2$ at zero magnetic field \cite{goldman_zero-field_2023,dong_composite_2023}. This state, dubbed ``anomalous composite Fermi liquid'' (ACFL),  arises from strong electron-electron interaction in a partially filled Chern band and serves as the parent state of the   fractional quantum anomalous Hall states observed in $t$MoTe$_2$ and rhombohedral graphene moir\'e superlattice  at Jain sequence filling factors $\nu=n/(2n\pm 1)$ \cite{cai2023signatures,zeng2023thermodynamic,park2023observation,
xu_observation_2023,lu_fractional_2023}. We show that the ACFL represents a novel compressible phase of matter with vanishing Drude weight, and is therefore fundamentally distinct from the Fermi liquid (FL) state which has a finite Drude weight.

{\it Compressible states in Chern band}. 
Generally speaking, the ground state of partially-filled Chern bands depends on the bandwidth, the band gap to neighboring bands, the band wavefunctions and electron-electron 
interactions. Large experimental and theoretical efforts have focused on studying the possible incompressible states that could arise in moir\'e Chern bands, especially when interactions are strong. Notable examples include fractional Chern insulators, charge density waves  and various other ordered states \cite{abouelkomsan2020particle,repellin2020chern,ledwith2020fractional,xie2021fractional,liu_recent_2024,crepel2023anomalous,li2021spontaneous,reddy2023fractional,wang2023fractional,sheng2024quantum}. 

On the other hand, less is understood about compressible states in partially filled Chern bands. The most familiar one is the Fermi liquid (FL) which is formed when the bandwidth is very large compared to the interaction strength or, even more exotically, could arise due to strong interaction-induced kinetic energies \cite{abouelkomsan2023quantum,reddy2023toward,liu_broken_2024}.
In the opposite limit when the bandwidth is narrow, a composite Fermi liquid may be realized at even-denominator filling factors, which shares many similarities with the celebrated CFL in the lowest Landau level (LLL) \cite{halperin_theory_1993}. 
Given that both FL and CFL states can be realized in a Chern band system, a natural question is how these two compressible states can be distinguished.  As we shall show in this work, in the clean limit, the composite Fermi liquid and Fermi liquid are sharply distinguished by the Drude weight.



For concreteness, we consider Chern bands in twisted semiconductor bilayers $t$MoTe$_2$ and $t$WSe$_2$, described by a two-layer continuum model \cite{wu_topological_2019, devakul2021magic}. Under certain conditions, this model can be approximately mapped onto Landau levels subject to a periodic potential \cite{morales2023magic, paul2023giant}. Motivated by this mapping, 
we study a periodically modulated lowest Landau level \cite{thouless1982quantized, usov1988theory,pfannkuche_theory_1992,paul_magic2022} as the simplest case of a Chern band. Depending on its bandwidth, which is controlled with the periodic potential, the system at half filling can be tuned from the composite Fermi liquid to the Fermi liquid, allowing a systematic study for the two gapless phases. 

We consider a periodic potential taking the form $V(\mathbf{r}) = \sum_{\mathbf{g}} V_{\mathbf{g}} e^{i \mathbf{g} \cdot \mathbf{r}}$ where $\{\mathbf{g}\}$ are the relevant harmonics. 
We focus on the simplest case where the unit cell of the periodic potential encloses one flux quanta, so that the Landau level is broadened by the potential but does not split. 
We further assume that the harmonics $|V_{\mathbf{g}}|$ are small relative to the cyclotron gap ($|V_{\mathbf{g}}|/\hbar \omega_c \ll 1$) so it doesn't mix different Landau levels. Under these assumptions, the $n$th flat Landau level  acquires an energy-momentum dispersion \cite{pfannkuche_theory_1992,supp} $E_n(\mathbf{k}) = \hbar \omega_c (n+\frac{1}{2}) + \tilde{E}_n(\mathbf{k})$ with  \begin{equation}
\label{eq:dispersion}
    \tilde{E}_n(\mathbf{k}) =  \sum_{\mathbf{g}}  V_{\mathbf{g}} \eta_{\mathbf{g}} e^{-il^2(\mathbf{g}\wedge\mathbf{k})} e^{-l^2|\mathbf{g}|^2/4}  L_n(\frac{l^2 |\mathbf{g}|^2}{2})
\end{equation}
where $L_n(\mathbf{q})$ is the $n$th Laguerre polynomial and $l = \sqrt{\hbar/eB}$ is the magnetic length. $\eta_{\mathbf{g}}$ is $1$ if $\mathbf{g}/2$ is an allowed reciprocal lattice vector and $-1$ otherwise.

If interactions are also much weaker than the band gap, we are justified to focus on the interacting problem within a single band neglecting band mixing. Taking the the LLL ($n = 0$), this gives rise to the following Hamiltonian
\begin{equation}
\label{eq:Ham_perturb}
\begin{aligned}
        & H = \sum_{\mathbf{k}} \tilde{E}_0(\mathbf{k})c^{\dagger}_{\mathbf{k}}c_{\mathbf{k}} + \\  & \sum_{\mathbf{k}_1,\mathbf{k}_2,\mathbf{q}} U_{\mathbf{q}}e^{-l^2|\mathbf{q}|^2/2} \lambda(\mathbf{k}_1,\mathbf{q}) \lambda(\mathbf{k}_2,-\mathbf{q})c^{\dagger}_{\mathbf{k}_1+\mathbf{q}}c^{\dagger}_{\mathbf{k}_2-\mathbf{q}} c_{\mathbf{k}_2} c_{\mathbf{k}_1}  
\end{aligned}
\end{equation} 
where $c^{\dagger}_{\mathbf{k}}$ is the lowest Landau level (LLL) creation operator with momentum $\mathbf{k}$ and we have dropped the constant factor $\frac{1}{2}\hbar \omega_c$. $\lambda(\mathbf{k},\mathbf{q})$ is the form factor of the LLL \cite{supp}. In what follows, we consider a two-dimensional long range Coulomb interaction $U_{\mathbf{q}} = \frac{U_0 a_0}{2 S}\frac{2\pi}{|\mathbf{q}|}$ where $U_0 = \frac{e^2}{4 \pi \epsilon \epsilon_0  a_0}$ is the strength of the interaction, $a_0$ is the period of the potential and $S$ is the area of the system.

The Hamiltonian \eqref{eq:Ham_perturb} defines a Chern band system. 
Without the periodic potential ($V_{\mathbf{g}} = 0$), it reduces back to the LLL problem. At half filling in such a case, the interacting ($U_0 \neq 0$) many-body ground state is the composite Fermi liquid \cite{halperin_theory_1993}. 
The composite fermion is constructed by attaching $2$ flux quanta to each electron, which experiences an effective magnetic field $B^\ast = B - 2 n \phi_0$ that is reduced from the external field \cite{jain2007composite}. 
At half filling $\nu = 1/2$, the effective magnetic field seen by the CFs vanishes and the resulting state is a Fermi liquid of composite fermions (CFL). 


In the opposite limit when interactions are turned off ($U_0$ = 0) and only the periodic potential is present ($V_{\mathbf{g}} \neq 0$), the ground state at partial filling is a normal Fermi liquid (FL). The model \eqref{eq:Ham_perturb} is therefore the minimal model to study phase transitions between the FL and CFL states as a function of the periodic potential strength relative to Coulomb interaction.

\textit{Drude Weight}. 
In general, the longitudinal conductivity of a clean system  at low frequency can have a singular part of the form $\sigma_{xx}(\omega)= D (\delta(\omega) + \frac{i}{\pi \omega} )$, where the coefficient of the delta function peak $D$ is known as the Drude weight \cite{kohn_theory_1964}. $D$ can be extracted from the imaginary (non-dissipative) part of $\sigma_{xx}$: $D = \pi \text{lim}_{\omega \rightarrow 0} \> \>  \omega \> \text{Im} \> \sigma(\omega)$. 
The Drude weight is responsible for optical absorption at low frequency. Gapless metallic states such as Fermi liquids generally have a non-zero Drude weight $D \neq 0$, while gapped insulating states with a finite spectral gap have zero Drude weight $D  = 0$ \cite{scalapino_insulator_1993}. 

The absence of a charge gap implies finite compressibility $\kappa = \frac{\partial n} {\partial \mu}$ with $\mu$ the chemical potential. In a clean system, one might expect that the presence of gapless low-energy charge excitations generally leads to finite Drude weight.
However, as we shall see, the composite Fermi liquid state in Chern bands is a compressible state of matter ($\mathbf{\kappa } \neq 0$) but has zero Drude weight ($D = 0$).

Note that in the absence of periodic potential, our Chern band reduces to a Landau level in a Galilean invariant system, where the presence of Galilean invariance dictates the longitudinal conductivity to be of the form  $\sigma_{\rm LL}(\omega) \propto \frac{n}{m} (\delta(\omega - \omega_c) + \delta(\omega + \omega _c))$ where $\omega_c = eB/m $ is the cyclotron frequency. Thus, the absence of optical absorption at low frequency $\omega < \omega_c$ implies vanishing Drude weight for all ground states in Landau Levels of Galilean invariant systems \cite{kohn_cyclotron_1961}, including the CFL state at half filling. In contrast, generic Chern band systems lack Galilean invariance, which allows finite Drude weight in principle.  



As realized in the seminal work by Kohn \cite{kohn_theory_1964},  the Drude weight for generic interacting Hamiltonians can be expressed as the second derivative of the ground state energy with respect to twisted boundary conditions  such that the wavefunction satisfies 
$\psi(\mathbf{r} + \mathbf{L})= e^{i \phi} \psi(\mathbf{r})$ for $\phi \in [0,2\pi)$,  which is equivalent to threading a flux through the cycles of the system. The Drude weight is given by (in units of $\hbar = 1 $)

\begin{equation}
\label{eq:drudeweight}
D(\phi) = \pi e^2 \dfrac{\partial^2 E_G(\phi)}{\partial \phi^2}.
\end{equation}
It's crucial that the derivative with respect to the flux in \eqref{eq:drudeweight} is evaluated first  before the thermodynamic limit \cite{scalapino_insulator_1993}.

In practice, there exist multiple methods for computing the Drude weight in finite systems (which become equivalent in the thermodynamic limit) \cite{koretsune_exact_2007}. A straightforward approach is to evaluate the Drude weight at a fixed value of the inserted flux $\phi = \phi_0$ and then extrapolate to the thermodynamic limit. 
Alternatively,  one may evaluate the Drude weight at flux $\phi = \phi_\text{min}$ corresponding to where the ground state energy is minimized. For this method, $\phi_\text{min}$ generally depends on the system size 
which makes finite-size scaling challenging. 

To mitigate the finite-size effect, we choose to average over different boundary conditions. This has been shown to yield excellent results  in small systems \cite{poilblanc_twisted_1991,koretsune_exact_2007}. We compute the Drude weight by averaging over all fluxes in one direction, ${\langle D\rangle} = \frac{1}{2 \pi}\int^{2\pi}_0 d\phi D(\phi)$. It is important to note that by definition, the average Drude weight ${\langle D\rangle}$ vanishes when the ground state energy $E_G(\phi)$ changes smoothly with the flux. The average ${\langle D\rangle}$ can be nonzero only when unavoided level crossings with excited states occur in the energy spectrum $E(\phi)$. Indeed, this is the case for non-interacting metals, where level crossings between single-particle states of different momenta are induced upon varying the flux or equivalently shifting the single-particle momenta $\mathbf{k} \rightarrow \mathbf{k} + \phi/L $. 

Generally speaking, when level crossings are present, $E_G(\phi)$ shows kinks as a function of $\phi$, thus allowing for a nonzero Drude weight ${\langle D\rangle}$.     
Specifically, at each level crossing, the 
first derivative of energy with respect to the flux is discontinuous, thus giving rise to a finite value of the integral ${\langle D\rangle} \propto \sum_{p}(\frac{\partial E_0}{\partial \phi}|_{p_{+}} - \frac{\partial E_0}{\partial \phi}|_{p_{-}}) $ where $p$ runs over all level crossings and $\pm$ refers to evaluating the first derivative from the left ($+$) or the right ($-$).

\begin{figure}[t!]
    \centering
    \includegraphics[width =0.7 \linewidth]{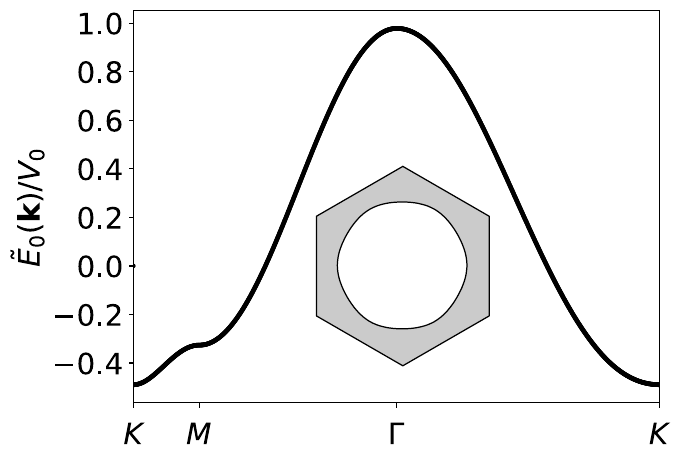}
    \caption{The energy dispersion $\tilde{E}_{0}(\mathbf{k})$ (equation \eqref{eq:dispersion}) along a cut in the Brillouin zone. The inset shows the Fermi surface contours at half filling where the shaded region denotes occupied states in the non-interacting limit. }
    \label{fig:n_one1}
\end{figure}

In the absence of the periodic potential, our system reduces to the LLL. Here, inserting the flux along the $x$ direction simply translates the Landau orbitals in the Landau gauge along $y$, which does not affect the ground state energy. Thus, it follows from Kohn's formula \eqref{eq:drudeweight} that the Drude weight is zero. 
We now study the evolution of the Drude weight as the periodic potential strength increases. 
For a weak potential, the first order correction to the many-body ground state energy $E_G$ is given by 
$\delta E = \bra{G}V\ket{G} = \int  d \mathbf{r} \rho(\mathbf{r}) V(\mathbf{r})$.     
Here $\ket{G}$ is the unperturbed CFL ground state in the half-filled LLL in the absence of the periodic potential, which has a uniform density $\rho(\mathbf{r})=\rho_0$. 
This implies that $\delta E = 0$, for all inserted flux. Therefore, the Drude weight  \eqref{eq:drudeweight} remains zero to first order in the periodic potential strength $V_0$. 

This contrasts sharply with the noninteracting case ($U_0=0$) where the ground state at partial filling is a noninteracting band metal with \textit{finite} Drude weight \cite{andjelkovic2018dc,supp} given by 


\begin{eqnarray}
\label{eq:nonintD}
D^{0}_{\alpha \beta} =  \pi e^2  \int_{\rm BZ} \frac{d^2 k}{(2 \pi)^2} \theta(\mu - \tilde{E}_0(\mathbf{k})) m^{-1}_{\alpha \beta}(\mathbf{k})     
\end{eqnarray} 
with $m^{-1}_{\alpha \beta}(\mathbf{k}) = \frac{1}{\hbar^2}\frac{\partial^2 \tilde{E}_0(\mathbf{k})}{\partial k_\alpha k_\beta} $ the inverse effective mass 
and $\alpha,\beta = x,y$. The integral runs over the Brillouin zone, $\theta$ is a step function, and $\mu$ is the chemical potential. Since the band dispersion $\tilde{E}_0(\mathbf{k})$ is induced by the periodic potential, the Drude weight in the noninteracting limit appears at first order in $V_0$.  

\begin{figure}[t!]
    \centering
    \includegraphics[width =\linewidth]{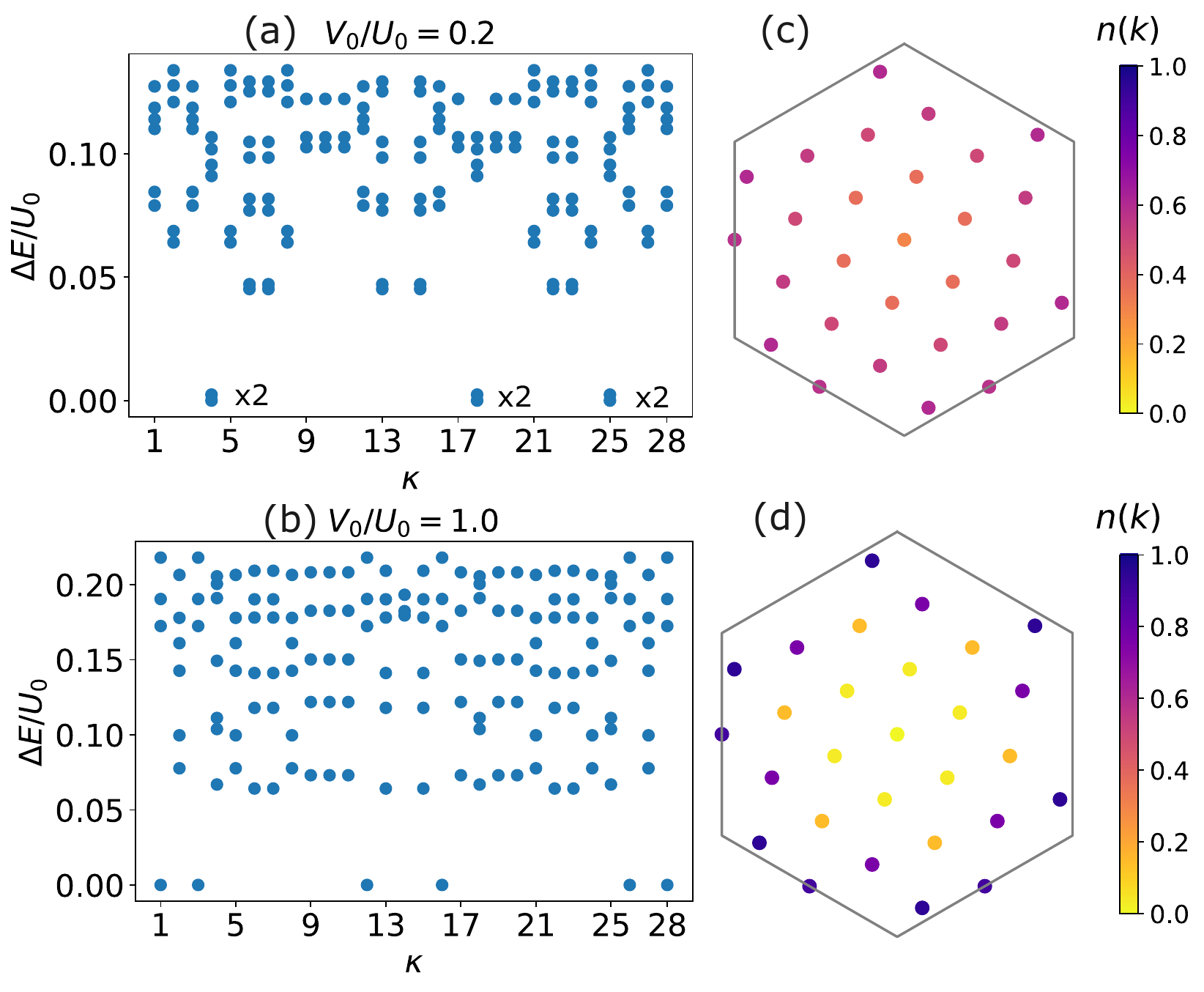}
    \caption{(a)-(b) The many-body spectrum in the CFL phase (a) and the FL phase (b). (c)-(d) Momentum space occupation number $n(\mathbf{k}) = \langle c^{\dagger}_{\mathbf{k}} c_{\mathbf{k}}\rangle $ of the many-body ground states in (a) the CFL phase and (b) the FL phase. $V_0/U_0$ denotes the ratio of the periodic potential strength to the interaction strength. Calculations are done at zero flux $\phi =0 $. $\kappa$ labels the total momentum sectors as defined in the supplementary material \cite{supp}.}
    \label{fig:n_one2}
\end{figure}

To study an arbitrary ratio of periodic potential and interaction strengths, we numerically solve the Hamiltonian \eqref{eq:Ham_perturb}
using exact diagonalization \cite{supp}. For concreteness, we consider first a $C_6$ symmetric potential with the lowest harmonics only, $|V_{\mathbf{g}}| = - V_0 $ for $\mathbf{g}_i= \frac{4 \pi}{\sqrt{3} a_0} (\cos 2\pi (i-1)/3,\sin 2\pi (i-1)/3)$ for $ i = 1,\dots,6$. $a_0$ is the potential period, chosen such that the unit cell of the potential encompasses $2 \pi$ flux.  In Fig \ref{fig:n_one1}, we plot the single-particle energy dispersion $\tilde{E}_{0}(\mathbf{k})$ given by Eq.\eqref{eq:dispersion}.

For weak to moderate values of  the periodic potential ($V_0/U_0 \leq 0.6$), we find the system at half filling to be a CFL. In Fig \ref{fig:n_one2}(a), we show the many-body energy spectrum for $V_0/U_0 = 0.2$. 
The structure of the low-lying states follows momentum configurations which correspond to the most compact Fermi sea of composite fermions \cite{rezayi_fermi-liquid-like_1994,rezayi_incompressible_2000,supp}. 
Moreover,  each momentum sector hosts a pair of nearly-degenerate ground states \cite{goldman_zero-field_2023, dong_composite_2023}, reminiscent of the center-of-mass degeneracy of the CFL state in a perfectly clean LLL.  
When the strength of the periodic potential increases, the system undergoes a transition to a normal Fermi liquid phase. In Fig \ref{fig:n_one2}(b), the many-body spectrum at $V_0/U_0 = 1$ clearly shows a distinct set of low-lying states,  which correspond to shell filling of the single-particle energy levels $\tilde{E}_{0}(\mathbf{k})$. Further evidence comes from momentum space occupation of the ground states which shows a well-defined Fermi surface, in sharp contrast to the CFL phase (Fig \ref{fig:n_one2}(c)) where the momentum space occupation is relatively uniform. 

\begin{figure}[t!]
    \centering
    \includegraphics[width =  \linewidth]{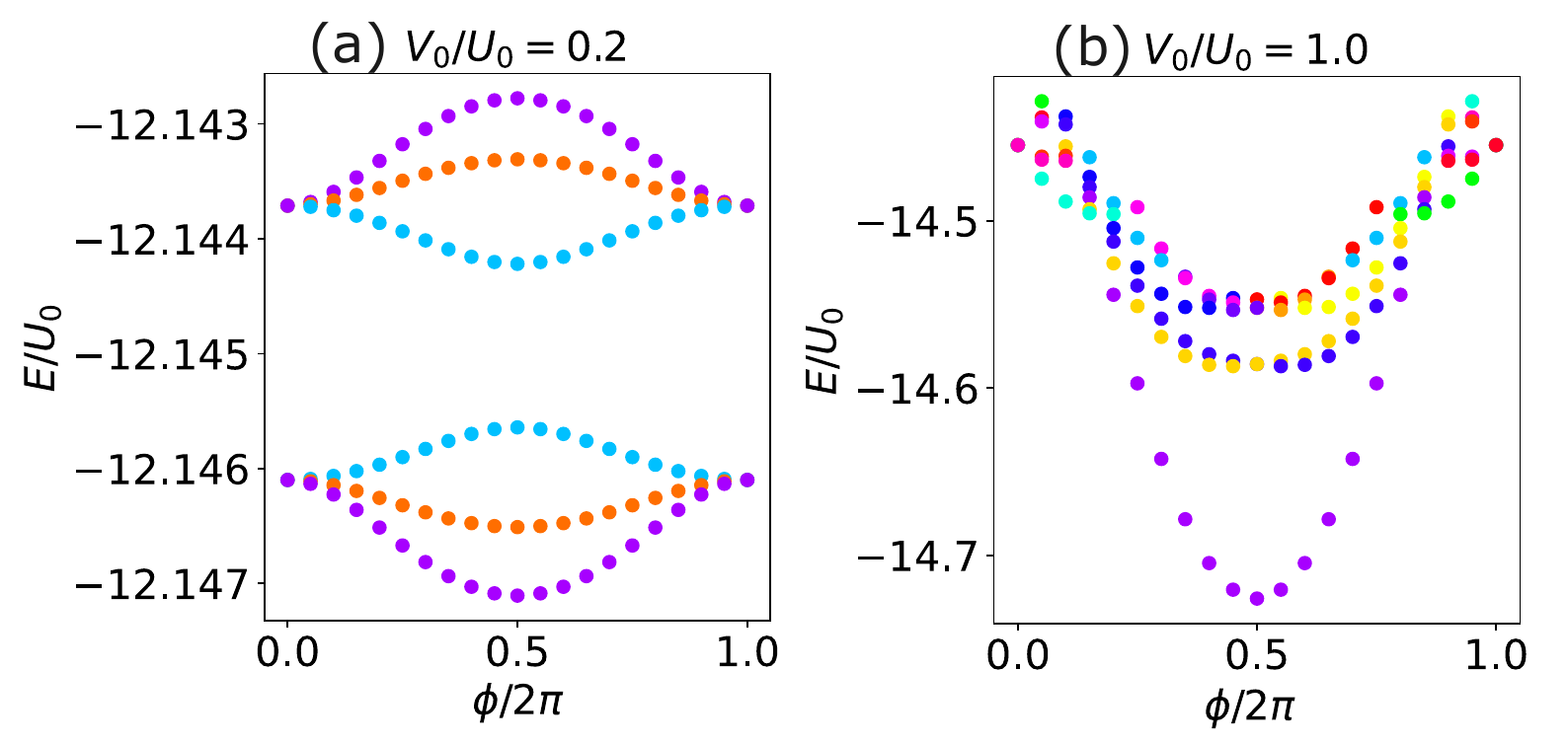}
    \caption{Spectral flow of the low-lying energy states in (a) the composite Fermi liquid phase and (b) the Fermi liquid phase. The colors denote different total momentum sectors. A flux is inserted in one of the two identical cycles of the cluster \cite{supp}. }
    \label{fig:n_one3}
\end{figure}

The distinction between both CFL and FL phases is further reflected on the spectral flow of the many-body ground states. While both phases are gapless in nature, we find very contrasting behavior under flux insertion. In a normal Fermi liquid, low-energy excitations live on a sharply defined Fermi surface in momentum space. Flux insertion amounts to shifting the single particle momenta and results in multiple level crossings between different momentum sectors as a function of the flux,    
as evident from Fig \ref{fig:n_one3}(b). Correspondingly, 
the lowest energy ground state changes its momentum sector, leading to kinks in $E_G(\phi)$.

In contrast, the spectrum of CFL phase exhibits a smooth spectral flow as evident in Fig \ref{fig:n_one3}(a). The degenerate ground states in the different momentum sectors vary smoothly with the flux. 
Remarkably, in the CFL phase we find that  the lowest energy state on the 28-site cluster stays in the same momentum sector throughout the full range of the flux. 
In other words,  
there are no level crossings or kinks in $E_G(\phi)$ \footnote{It's important to note that level crossings can exist independent of the underlying phase. Indeed, the total momentum changes as $\mathbf{K} \rightarrow \mathbf{K} + N_p \phi/L $ when flux $\phi $ is inserted where $N_p$ is the number of particles. 
Depending on the system geometry, the ground state total momenta can flow from one sector to another. However, level crossings in the FL phase occur at generic flux values not dictated by this relation while this is not the case for the CFL ground states.}.

\begin{figure}[t!]
    \centering
    \includegraphics[width = 0.9\linewidth]{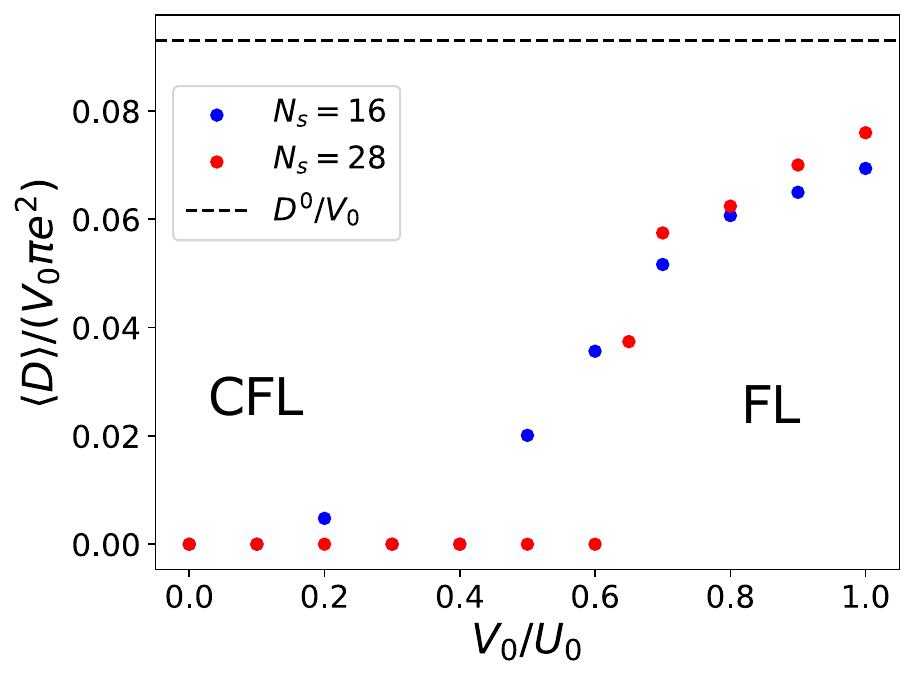}
    \caption{Drude weight (in units of $V_0$) averaged over twisted boundary conditions, ${\langle D\rangle} = \frac{1}{2 \pi} \int_0^{2\pi} D(\phi) d\phi $ as a function of the ratio of the periodic potential strength to the interaction strength $V_0/U_0$. $D^0 = (D^{0}_{xx} + D^{0}_{yy})/2  $ is the non-interacting ($U_0 = 0$) Drude weight in the thermodynamic limit obtained from equation \eqref{eq:nonintD}.
    $N_s$ denotes the number of unit cells \cite{supp}.}
    \label{fig:n_one4}
\end{figure}

We further calculate the Drude weight as a function of the ratio of periodic potential strength to the interaction strength $V_0/U_0$. 
In Fig \ref{fig:n_one4}, we plot the Drude weight calculated using the averaging method. In the CFL regime (approximately $V_0/U_0 \leq 0.6$), we find the Drude weight to (nearly) vanish. Upon increasing $V_0/U_0$, we notice increasing values of the Drude weight as the system undergoes the transition to the Fermi liquid regime where the Drude weight is non-zero.
While averaging over boundary conditions clearly illustrates the difference between the two phases, our conclusions remain the same if instead the Drude weight is calculated without averaging \cite{supp}. 

Interestingly, our numerical study of the Drude weight, shown in Fig. \ref{fig:n_one4}, suggests a direct transition between CFL and FL phases as a function of $V_0/U_0$.  
Determining the nature of this phase transition \cite{barkeshli_continuous_2012,song_phase2024} from finite systems studies can be challenging and requires careful analysis which we defer to the future.

\begin{figure}[t!]
    \centering
    \includegraphics[width = \linewidth]{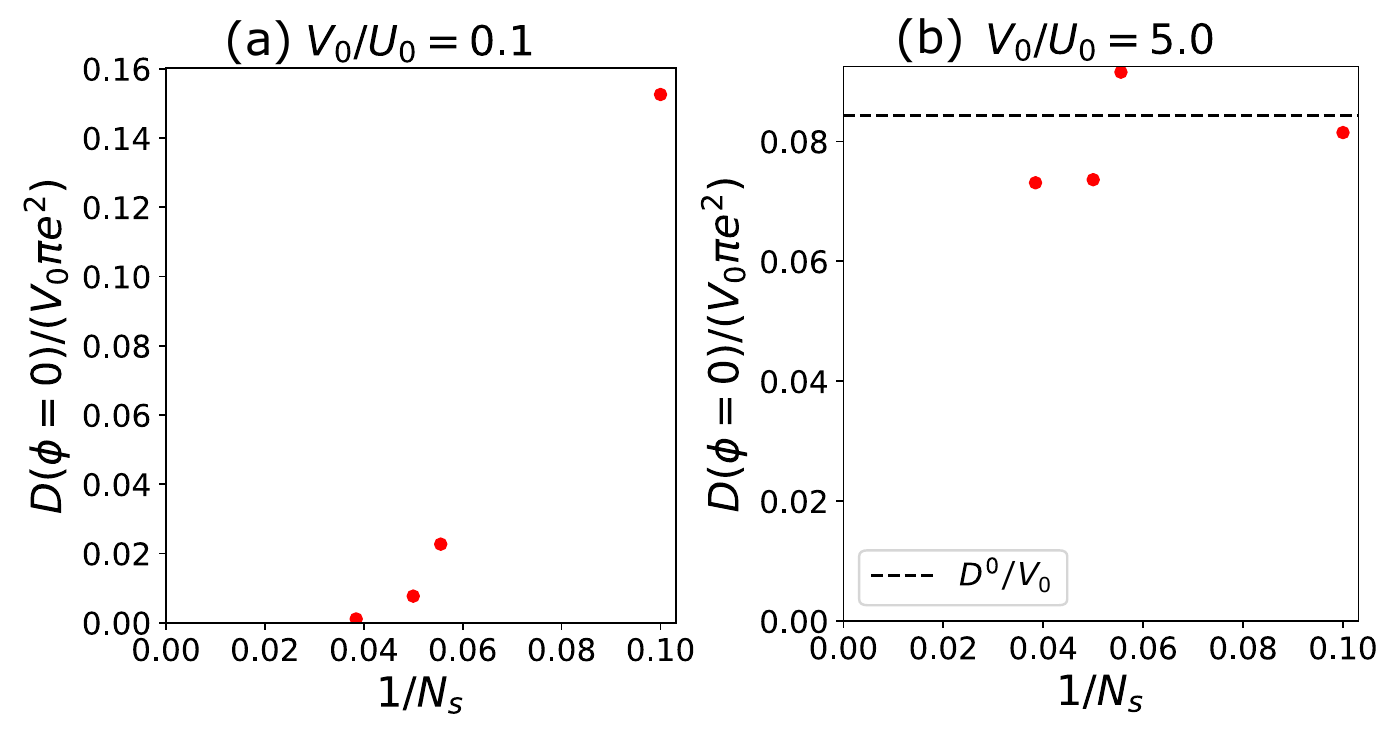}
    \caption{Drude weight \eqref{eq:drudeweight} (in units of $V_0$)  calculated at zero flux $\phi = 0$ for the LLL modulated with $C_4$ symmetric potential in the CFL phase (a) and the FL phase (b). $D^0 = (D^{0}_{xx} + D^{0}_{yy})/2  $ is the non-interacting ($U_0 = 0$) Drude weight in the thermodynamic limit obtained from equation \eqref{eq:nonintD}. $N_s$ denotes the number of unit cells \cite{supp}. }
    \label{fig:n_one5}
\end{figure}

Lastly, we demonstrate the generality of our conclusions by considering another example: interacting electrons in the lowest Landau level with a $C_4$ symmetric periodic potential of the form $V(\mathbf{r}) = 2 V_0 [\cos(\frac{2 \pi}{a_0} r_x)  + \cos(\frac{2 \pi}{a_0} r_y)]$ with the period $a_0$  chosen such that the unit cell encloses one flux quantum \cite{supp}. Similar to the previous case, when the ratio $V_0/U_0$ is small, the ground state is a CFL \cite{supp}. In Fig. \ref{fig:n_one5}(a), we plot the Drude weight \eqref{eq:drudeweight} (obtained without averaging) for different system sizes and find that it monotonically decreases with increasing system size suggesting that it vanishes in the thermodynamic limit. In contrast, we find the Drude weight in the FL phase to be clearly finite as evident from Fig. (\ref{fig:n_one5}(b)). This further supports our conclusion and shows that the vanishing of the Drude weight in the CFL is a universal feature independent of the underlying model.

\textit{Discussion.}
In this paper, we have shown that composite Fermi liquids in Chern bands have zero Drude weight. By introducing a minimal model consisting of Landau levels with a periodic potential, Galilean invariance is broken, but the vanishing Drude weight still persists in the CFL regime. We have provided evidence for such a conclusion using a perturbative argument when the potential strength is weak combined with numerical results from exact diagonalization. Even in systems where the Chern band wavefunctions are different from those of the LLL (such as twisted $\text{MoTe}_2$), we have also observed a vanishing Drude weight in the CFL regime \cite{supp}.


We have established that the absence of Drude weight is the defining property of the composite Fermi liquid as a unique gapless phase distinct from other types of non-Fermi liquids (which by definition have vanishing quasiparticle weight).   
Our work therefore shows that the composite Fermi liquid state can be experimentally established by measuring electronic compressibility and low-frequency optical conductivity.


We note that the absence of Drude weight is consistent with low-frequency conductivity of the form $\sigma_{xx} \propto i\omega$ (instead of $1/i\omega$) suggested by the Ioffe-Larkin rule for the  composite Fermi liquid in Chern bands as shown in a semiclassical treatment in Ref. \cite{stern_transport_2023}, implying that $\sigma_{xx}(\omega \rightarrow 0) = 0$.  We leave the explicit calculation of the full optical spectrum of CFL to future work. 



While the CFL phase breaks time reversal symmetry either through an external magnetic field or spontaneously \cite{goldman_zero-field_2023, dong_composite_2023}, an intriguing possibility is the realization of two copies of a CFL related by time-reversal symmetry \cite{myerson2023conjugate,shi2024excitonic} which could be realizable in moir\'e systems having a time-reversed pair of Chern bands. In the absence of any interaction between the two flavors, the Drude weight of this time-reversal-symmetric phase vanishes in a similar manner to a single CFL as shown above. 
We leave the detailed study of the full problem to future work.

Finally, we have neglected disorder in our analysis. It is known \cite{halperin_theory_1993} that for the CFL in the LLL, the DC longitudinal conductivity is linear in the wavevector. Disorder can mix different wavevectors such that the DC conductivity becomes finite. The same reasoning applies to generic CFL states in Chern bands: such Drudeless metal acquires DC longitudinal conductivity in the presence of disorder, which is inversely proportional to the mean free path.


\begin{acknowledgements}
We thank Aidan Reddy for helpful discussions. This work was supported by  the Air Force Office of Scientific Research (AFOSR) under Award No. FA9550-22-1-0432 and benefited from computing resources provided by the MIT SuperCloud and Lincoln Laboratory Supercomputing Center. A.A was supported by the Knut and Alice Wallenberg Foundation (KAW 2022.0348). N.P. acknowledges the KITP graduate fellowship. 
L. F. was partly supported by the Simons Investigator Award from the Simons Foundation. A.S. was supported by grants
from the ERC under the European Union’s Horizon 2020
research and innovation programme (Grant Agreements
LEGOTOP No. 788715 ) and the DFG (CRC/Transregio
183, EI 519/71).
\end{acknowledgements}

\bibliography{refs}

\appendix

 \begin{widetext}
      		\renewcommand{\theequation}{S\arabic{equation}}
 		\setcounter{equation}{0}
 		\renewcommand{\thefigure}{S\arabic{figure}}
 		\setcounter{figure}{0}
 		\renewcommand{\thetable}{S\arabic{table}}
 		\setcounter{table}{0}

\section{Supplemental Material for Compressible quantum liquid with vanishing Drude weight}

In this supplemental material, we elaborate on the models used in the main text, provide additional results and explain the exact diagonalization methods.

\section{Periodically modulated Landau levels}
In this section, we introduce the model studied in the main text which consists of Landau levels in addition to a periodic potential. Due to magnetic translation symmetry of the quantum Hall problem, it is possible to choose a unit cell defined by two basis vectors $\mathbf{a}_1$ and $\mathbf{a}_2$ that encloses $ 2 \pi$ flux, $ |\mathbf{a}_1 \wedge \mathbf{a}_2| = 2 \pi l^2$  where $l  = \sqrt{\hbar /eB}$ is the magnetic length. Such a choice is not unique and any $2 \pi$ flux-enclosing unit cell describes the problem equivalently. One then obtains Bloch-like wavefunctions as representations of the magnetic translation symmetry algebra. Let us denote them by $\ket{n,\mathbf{k}}$ where $n$ is the LL index and $\mathbf{k}$ is the "Bloch" momentum. In the symmetric gauge, these wavefunctions are known to be the modified Weierstrass sigma functions \cite{haldane_modular-invariant_2018,ferrari_two-dimensional_1990,ferrari_wannier_1995,wang_exact_2021}

We consider adding a periodic potential of the form $V(\mathbf{r}) = \sum_{\mathbf{g}} V_{\mathbf{g}} e^{i \mathbf{g} \cdot \mathbf{r}}$ where $\{\mathbf{g}\}$ are the relevant harmonics which set the periodicity of the potential.  Let's denote the two basis vectors that define the unit cell of the potential to be $\mathbf{R}_1$ and $\mathbf{R}_2$ such that $V(\mathbf{r}  + n \mathbf{R}_1 + m \mathbf{R}_2) = V(\mathbf{r})$ for integer $n$ and $m$. The reciprocal basis vectors are obtained from the relation , $\mathbf{R}_i \cdot \mathbf{g}_j = 2 \pi \delta{_{ij}}$. The harmonics of the potential can be then expanded in these basis, $\mathbf{g} = n \mathbf{g}_1 + m \mathbf{g}_2$. We consider the case where the unit cell of the periodic potential encloses $2 \pi $ flux quanta, $|\mathbf{R}_1 \wedge \mathbf{R}_2| = 2 \pi l^2 $. Since the choice of the quantum Hall basis vectors $\mathbf{a}_1$ and $\mathbf{a}_2$ is not unique, we can always take $\mathbf{a}_1 = \mathbf{R}_1$ and $\mathbf{a}_2 = \mathbf{R}_2$. The full Hamiltonian is given by,

\textcolor{black}{\begin{equation}
\label{eq:Ham_supp}
\begin{aligned}
    H = \sum_{\mathbf{k}} \hbar \omega_c (n+\frac{1}{2}) c^{\dagger}_{n\mathbf{k}} c_{n\mathbf{k}} + \sum_{\mathbf{k}_1 \mathbf{k}_2 n_1 n_2} \sum_{\mathbf{g}} V_{\mathbf{g}} \braket{n_1,\mathbf{k}_1}{ e^{i \mathbf{g} \cdot \mathbf{r}}|n_2,\mathbf{k}_2} c^{\dagger}_{n_1 \mathbf{k}_1} c_{n_2 \mathbf{k}_2} \\ + \sum_{\mathbf{k}_1 \mathbf{k}_2 \mathbf{k}_3\mathbf{k}_4  \in {\rm BZ}} \sum_{\mathbf{q}} \sum_{n_1,n_2, n_3 ,n_4}  U_{\mathbf{q}}\braket{n_1,\mathbf{k}_1}{e^{i \mathbf{q} \cdot \mathbf{r}}|n_4,\mathbf{k}_4}
    \braket{n_2,\mathbf{k}_2}{e^{-i \mathbf{q} \cdot \mathbf{r}}|n_3,\mathbf{k}_3} c^{\dagger}_{n_1 \mathbf{k}_1} c^{\dagger}_{n_2 \mathbf{k}_2} c_{n_3 \mathbf{k}_3}  c_{n_4 \mathbf{k}_4} 
    \end{aligned}
\end{equation}}
\textcolor{black}{The first two terms correspond to the non-interacting part of the Hamiltonian while the third term corresponds to the Hamiltonian of two-body interactions $U(\mathbf{r}_1 - \mathbf{r}_2) = \sum_{\mathbf{q}} U_{\mathbf{q}} e^{i \mathbf{q} \cdot (\mathbf{r}_1 - \mathbf{r}_2)}$ written in the Landau level basis}. 

\textcolor{black}{ To evaluate the matrix elements $\braket{n_1,\mathbf{k}_1}{e^{i \mathbf{q} \cdot \mathbf{r}}|n_2,\mathbf{k}_2}$ for any momentum $\mathbf{q}$}  , we briefly review the algebra of the guiding center and relative coordinate operators. The position operator $\mathbf{r}$ can be decomposed into two parts $\mathbf{r} = \mathbf{R} + \bar{\mathbf{R}}$ with $\mathbf{R} = \mathbf{r} + (\hat{z} \wedge \boldsymbol{\Pi} ) l^2$  is the guiding center operator and $\bar{\mathbf{R}} = -(\hat{z} \wedge \boldsymbol{\Pi} ) l^2 $ is the relative (cyclotron) coordinate operator. $\boldsymbol{\Pi}$ denotes the canonical momentum operator , $\boldsymbol{\Pi} = - \boldsymbol{\partial} - e \mathbf{A}$ where $\mathbf{A}$ is the vector potential such that the magnetic field is  $B = \boldsymbol{\nabla} \wedge \mathbf{A}$. The guiding center and relatve coordinate operators satisfy the following commutation relations, $[R_a,R_b] = -i \epsilon_{ab} l^2$ , $[\bar{R}_a,\bar{R}_b] = i \epsilon_{ab} l^2$ and $[R_a,\bar{R}_b] = 0$. 

The guiding center operators are the generators of the magnetic translation algebra, $t(\mathbf{q}) = e^{i \mathbf{q}\cdot \mathbf{R}}$ obeying the following commutation relation \begin{equation}
\label{eq:boundary1}
    t(\mathbf{q}_1) t(\mathbf{q}_2) = e^{ i l^2 \mathbf{q}_1 \wedge \mathbf{q_2}} t(\mathbf{q}_2) t(\mathbf{q}_1)
\end{equation}

To satisfy the above algebra, the action of $t(\mathbf{q})$ on $\ket{n,\mathbf{k}}$ is given by \begin{equation}t(\mathbf{q}) \ket{n,\mathbf{k}} = e^{\frac{i}{2} l^2 \mathbf{q} \wedge \mathbf{k}} \ket{n,\mathbf{k} + \mathbf{q}} \end{equation} The wavefunction $\ket{n,\mathbf{k}}$ in the symmetric gauge satisfy the following Brillouin zone boundary condition \cite{wang_exact_2021} \begin{equation}
\label{eq:boundary2}
t(\mathbf{g}_i)
    \ket{n,\mathbf{k}} = (-1) e^{i l^2 \mathbf{g}_i \wedge \mathbf{k}} \ket{n,\mathbf{k}}
\end{equation} for $i = 1,2$. Combining \eqref{eq:boundary1} and \eqref{eq:boundary2}, we arrive to the following condition  \begin{equation}
\label{eq:boundary3}
    \ket{n,\mathbf{k} + \mathbf{g}_i} =  - e^{\frac{i}{2} l^2 \mathbf{g}_i \wedge \mathbf{k}}  \ket{n,\mathbf{k}}
\end{equation}
The relative coordinate operator $\bar{\mathbf{R}}$ moves a state different Landau levels. Explicitly, $\bar{R}_x = (\bar{a}^{\dagger} + \bar{a})/(\sqrt{2})$  and $\bar{R}_y = (\bar{a}^{\dagger} - \bar{a})/(\sqrt{2}i)$ where $\bar{a}^\dagger \ket{n,\mathbf{k}} = \sqrt{n+1} \ket{n+1,\mathbf{k}}$. Utilizing this and the condition \eqref{eq:boundary3}, we arrive at the following expression after lengthy algebra
\textcolor{black}{
\begin{equation}
\begin{aligned}
\braket{n_1,\mathbf{k}_1}{ e^{i \mathbf{q} \cdot \mathbf{r}}|n_2,\mathbf{k}_2} = G_{n_1 n_2}(\mathbf{q}) F({\mathbf{k}_1,\mathbf{k}_2},\mathbf{q}) 
\end{aligned}
\end{equation} 
where 
\begin{equation}
        G_{n_1,n_2}(\mathbf{q}) = e^{-l^2 |\mathbf{q}|^2/4} \sqrt{\dfrac{n_< !}{n_> !}} L_{n_<}^{|n_2-n_1|}\bigg(\dfrac{|\mathbf{q}|^2 l^2}{2}\bigg) \times  \begin{cases} 
      \big(\dfrac{-i l \bar{z}}{\sqrt{2}}\big)^{|n_2 - n_1|} & n_2 > n_1 \\
      \big(\dfrac{i l z}{\sqrt{2}}\big)^{|n_2 - n_1|} & n_1 \geq n_2
   \end{cases}
\end{equation} 
\begin{equation}
    F(\mathbf{k_1},\mathbf{k_2},\mathbf{q}) =  e^{-i l^2( (\mathbf{k}_2 + \mathbf{q}) \wedge \mathbf{k}_1 + \mathbf{q} \wedge \mathbf{k}_2}) \sum_{\mathbf{g}}  \eta_\mathbf{g} \delta_{\mathbf{k}_2 + \mathbf{q} -\mathbf{k}_1, \mathbf{g}}
\end{equation}}
where $n_> = {\rm max}(n_1,n_2)$ and $n_< = {\rm min}(n_1,n_2)$, $L_{a}^{b}(x)$ is the generalized Laguerre polynomial, $z = g_x + ig_y$ and $\eta_{\mathbf{g}} = 1 $ if $\mathbf{g}/2$ is allowed reciprocal lattice vector and $-1$ otherwise.

\textcolor{black}{Next, we assume both the periodic potential strength and the interaction strength are weak compared to the cyclotron frequency: $|V_{\mathbf{g}}| << \hbar \omega_c$, $|U_{\mathbf{q}}| << \hbar \omega_c$. We can then neglect any Landau level mixing in \eqref{eq:Ham_supp}. We discuss first the \textit{non-interacting part} of \eqref{eq:Ham_supp}. It describes \textit{dispersive} Landau levels with a dispersion $E_n(\mathbf{k}) = \hbar \omega_c (n+1/2) + \tilde{E}_n(\mathbf{k}) $ with $\tilde{E}_n(\mathbf{k}) = \sum_{\mathbf{g}} \braket{n,\mathbf{k}}{V_{\mathbf{g}} e^{i \mathbf{g} \cdot \mathbf{r}}|n,\mathbf{k} }$ given by equation (1) in the main text which we rewrite below }
\textcolor{black}{
\begin{equation}
    \label{eq:supp_disper}
    \tilde{E}_n(\mathbf{k}) =  \sum_{\mathbf{g}}  V_{\mathbf{g}} \eta_{\mathbf{g}} e^{-il^2(\mathbf{g}\wedge\mathbf{k})} e^{-l^2|\mathbf{g}|^2/4}  L_n(\frac{l^2 |\mathbf{g}|^2}{2})
\end{equation}
The \textit{interacting} part of \eqref{eq:Ham_supp} under this approximation describes projected interactions onto the $n$th LL. In our paper, we focus on the lowest Landau level ($n = 0$). This gives rise to the Hamiltonian (2) in the main text which we copy below}
\textcolor{black}{
\begin{equation}
\begin{aligned}
        & H = \sum_{\mathbf{k}} \tilde{E}_0(\mathbf{k})c^{\dagger}_{\mathbf{k}}c_{\mathbf{k}} +   \sum_{\mathbf{k}_1,\mathbf{k}_2,\mathbf{q}} U_{\mathbf{q}}e^{-l^2|\mathbf{q}|^2/2} \lambda(\mathbf{k}_1,\mathbf{q}) \lambda(\mathbf{k}_2,-\mathbf{q})c^{\dagger}_{\mathbf{k}_1+\mathbf{q}}c^{\dagger}_{\mathbf{k}_2-\mathbf{q}} c_{\mathbf{k}_2} c_{\mathbf{k}_1}  
\end{aligned}
\end{equation} 
where we have dropped the constant $ \frac{1}{2} \hbar \omega_c$ and the index $n = 0$ from the operators ${c_{\mathbf{k}}}$ for convenience and we define $\lambda(\mathbf{k},\mathbf{q}) \equiv F(\mathbf{k} +\mathbf{q},\mathbf{k},\mathbf{q})$.}

\section{Drude weight of Landau levels modulated with a periodic potential}

In this sections, we discuss the Drude weight in interacting Landau levels with periodic potential. The Drude weight can be generally be written \cite{Resta2018Sep} as  \begin{equation}\label{eq:Drude2}
    D_{\alpha \beta} = (ne^2/m)\delta_{\alpha \beta}+ S \sum_{n\neq m} f_{nm} \frac{[j_\alpha]_{nm} [j_\beta]_{mn}}{E_n-E_m}
\end{equation}
where $f_{nm} = f_n-f_m$ and $f_n = (e^{\beta(E_n-\mu)}+1)^{-1}$ , $j_{\alpha}$ is the current operator and $S$ is the area of the system. 

In the absence of a periodic potential, the problem reduces to the quantum Hall problem with a uniform magnetic field. The current operator is given by $\mathbf{j}  =  -e \vec{\Pi}/m$ and it has the matrix representation,
\begin{equation}
\label{eq:current_LL}
    [j_\alpha]_{cq;dq'} = e\delta_{qq'} (\sqrt{c+1}\delta_{c+1,d} + s \sqrt{c} \delta_{c-1,d})/(\sqrt{2s} \ell m S)
\end{equation}
where $s=\pm 1$ for $\alpha =x,y$ and $q$ is the intra-Landau level quantum number (e.g. momentum as defined in the previous section). \textcolor{black}{One immediately observes the current operator has only off-diagonal matrix elements in the inter-Landau level basis which means it only couples the $n$th Landau level with the the $n-1$ or $n+1$ Landau level with energy difference $\hbar \omega_c$. Therefore for any state, either interacting or non-interacting, we obtain the following  }
\begin{equation}
    \begin{aligned}
       \textcolor{black}{D_{\alpha \beta}} &= \frac{ne^2}{m} \delta_{\alpha\beta}+ \sum_{c\neq d,q,q'} f_{cd} \frac{[j_\alpha]_{cq;dq'} [j_\beta]_{dq';cq}}{E_c-E_d} \\
        &= \frac{ne^2}{m} \delta_{\alpha \beta}  +\frac{e^2}{2s\ell^2m^2} \delta_{\alpha \beta} \sum_{c\neq d,q,q'} f_{cd} \delta_{qq'}\frac{s d \delta_{c+1,d}+sc\delta_{c-1,d}}{E_c-E_d}\\
        &= \frac{ne^2}{m} \delta_{\alpha \beta}  +\frac{e^2}{2\ell^2m^2} \delta_{\alpha \beta}  \left(\sum_{d,q}f_{d-1,d}\frac{d}{-\omega_c} + \sum_{c,q} f_{c,c-1}\frac{c}{\omega_c} \right)\\
        &= (\frac{ne^2}{m} - \frac{ne^2}{m})\delta_{\alpha \beta}  = 0.
    \end{aligned}
\end{equation}
This is the familiar fact that longitudinal conductivity vanishes for Landau levels at any chemical potential or temperature. Indeed, this is a consequence of the $f$-sum rule and Kohn's theorem \cite{kohn_cyclotron_1961}, which dictates that the total optical absorption in Landau levels is saturated by the absorption peak at $\omega=\omega_c$.

We now turn to the case where there is a non-zero  periodic potential. \textcolor{black}{This will generally modify the current operator \eqref{eq:current_LL} and diagonal terms ($ c = d$ in equation \eqref{eq:current_LL}) can exist which means states within a single Landau level can be coupled by the current perturbation.} We have shown in the previous section that the addition of a periodic potential gives rise to a dispersion \eqref{eq:supp_disper}. 
  For simplicity, let's consider the lowest modulated Landau level at partial filling. In the absence of interactions, \textcolor{black}{ it's easier to calculate the \textit{projected} Drude weight from the formula below which is valid only for non-interacting cases, }
  
  \begin{eqnarray}
\label{eq:nonint_supp}
D_{\alpha \beta} = 2 \pi e^2  \int_{\rm BZ} \frac{d^2 k}{(2 \pi)^2} \theta(\mu - \tilde{E}_0(\mathbf{k})) m^{-1}_{\alpha \beta}(\mathbf{k})     
\end{eqnarray} 
with $m^{-1}_{\alpha \beta}(\mathbf{k})$ the inverse effective mass given by 
\begin{eqnarray}
m^{-1}_{\alpha \beta}(\mathbf{k}) = \frac{1}{\hbar^2}\frac{\partial^2 \tilde{E}_0(\mathbf{k})}{\partial k_\alpha k_\beta}
\end{eqnarray}
where $\alpha,\beta = x,y$. The integral runs over the Brillouin zone, $\theta$ is a step function, and $\mu$ is the chemical potential. Consider a periodic potential with $D_6$ periodicity that is a sum of the lowest harmonics . Using equation \eqref{eq:supp_disper}, this is simplified to \begin{equation}
\label{eq:dispersion_square}
    \tilde{E}_0(\mathbf{k}) = -V_0 e^{-l^2 |\mathbf{g}_0|^2/4} \sum_{i = 1,3,5} \cos( l^2 \mathbf{g}_i \wedge \mathbf{k})
\end{equation}
where $\mathbf{g}_i = |\mathbf{g}_0|(\cos[(i-1)\pi/3],\sin[(i-1)\pi/3])$ and $|\mathbf{g}_0| = 4 \pi/(\sqrt{3} a_0)$ with $a_0$ is the lattice constant of the periodic potential. The integral in \eqref{eq:nonint_supp} is generally non-zero unless the Landau level is fully filled therefore we conclude that the addition of a periodic potential that breaks Galilean invariance gives rise to a finite Drude weight in the non-interacting limit.

\textcolor{black}{In the presence of both interactions and periodic potential, the Drude weight can still be calculated using equation \eqref{eq:Drude2}. However, this equation requires the energies of all eigenstates. Instead we utilize the formula (equation 3 in the main text) derived by Kohn \cite{kohn_theory_1964} which only needs the ground state energy.}
\begin{figure}[t!]
    \centering
    \includegraphics[width = 0.8 \linewidth]{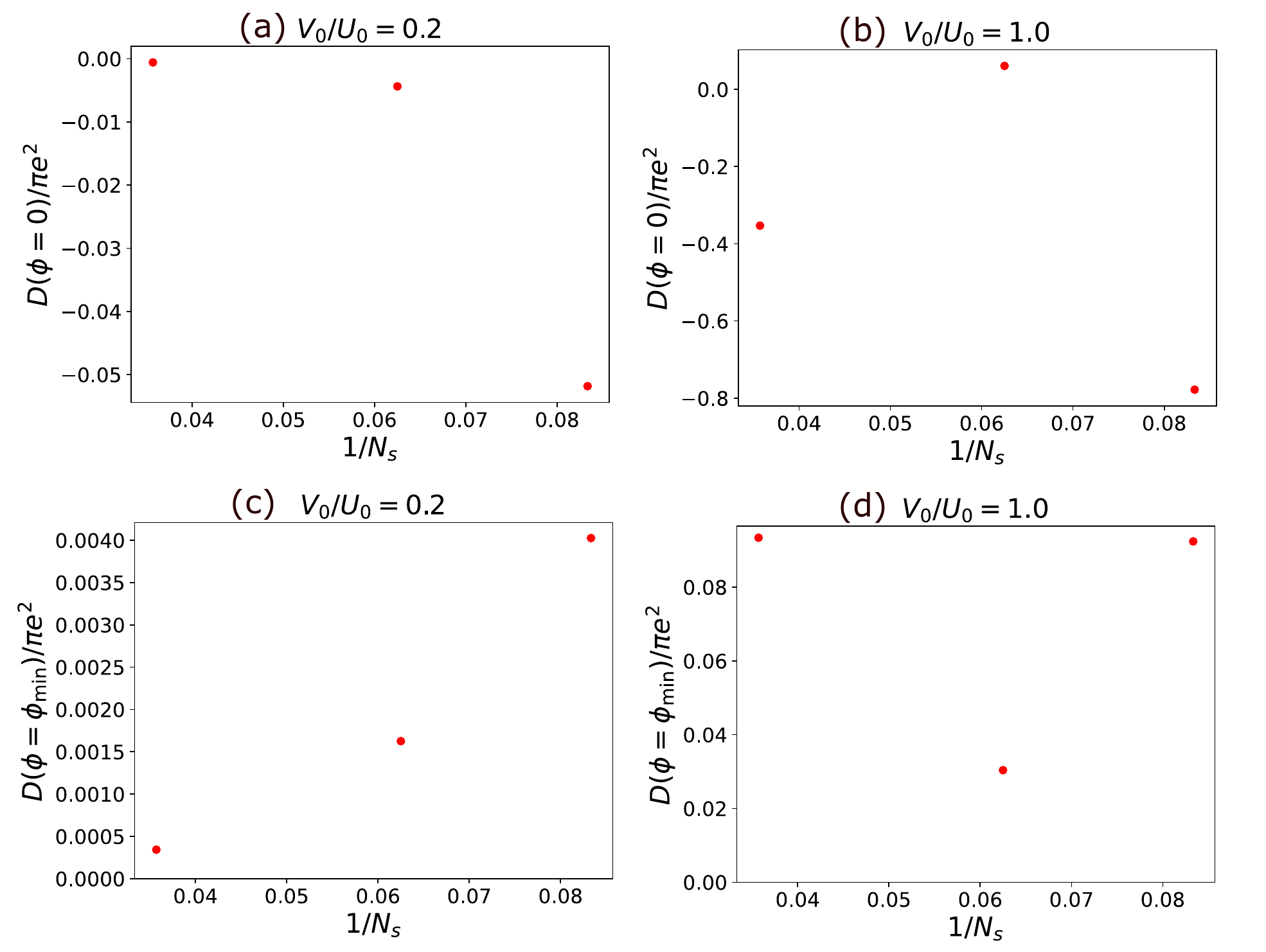}
    \caption{Drude weight for periodically modulated Landau Levels (equation (2) in the main text) evaluated at $\phi = 0.0$ in (a) and (b) and $\phi = \phi_{\rm min}$ in (c) and (d) corresponding to the flux that minimizes the energy. }
    \label{fig:S1}
\end{figure}
\section{Additional results for the Drude weight}
In the main text, we have calculated the Drude weight for interacting periodically modulated Landau levels with a $C_6$ symmetric potential using the averaging method (c.f Fig. 4. in the main text). Here we provide additional evidence that the Drude weight vanishes in the CFL phase. In Fig. \ref{fig:S1}, we show the Drude weight calculated at a zero flux $\phi = 0$ and at flux $\phi = \phi_{\rm min}$ which minimizes the ground state energy. In the CFL phase (for example when $V_0/U_0 = 0.2$), the Drude weight monotonically decreases as a function in system size as shown in Fig. \ref{fig:S1}(a) and Fig. \ref{fig:S1}(c) suggesting that it vanishes in the thermodynamic limit. In the Fermi liquid phase, we find the Drude weight to exhibit an opposite behavior as shown in Fig. \ref{fig:S1}(b) and Fig. \ref{fig:S1}(d) suggesting it remains finite in the thermodynamic limit. However, care should be exercised in interpolating due to the small number of data points (We only considered systems that have equal aspect ratio). The averaging method on the other hand mitigates all these subtleties and provide a clear approach to the thermodynamic limit.

\section{Drude weight for twisted $\text{MoTe}_2$}

In this section, we study the Drude weight for the recently predicted anomalous composite Fermi liquid phase in twisted $\rm{MoTe_2}$  \cite{goldman_zero-field_2023,dong_composite_2023}. We find a very similar behavior to the problem of periodically modulated Landau levels discussed in the main text. We focus on the first valence band with a Chern number $C = 1$ \cite{reddy2023fractional}. We consider Coulomb interaction of the form $V(\mathbf{q}) = \frac{e^2}{\epsilon \epsilon_0}\frac{2 \pi}{|\mathbf{q}|}$, projected onto this band and perform exact diagonalization at 
half filling. Previous studies \cite{goldman_zero-field_2023} have shown evidence for a composite Fermi liquid phase for $\epsilon = 10$. We find that the CFL phase is stabilized for even weaker interaction ($\epsilon^{-1} \approx 0.02$). As shown in Fig. \ref{fig:S_2}, we find the average Drude weight to vanish (Fig. \ref{fig:S_2}(b)) when in the system is in the composite Fermi liquid phase. 
\begin{figure}[t!]  
    \centering
    \includegraphics[width = \linewidth]{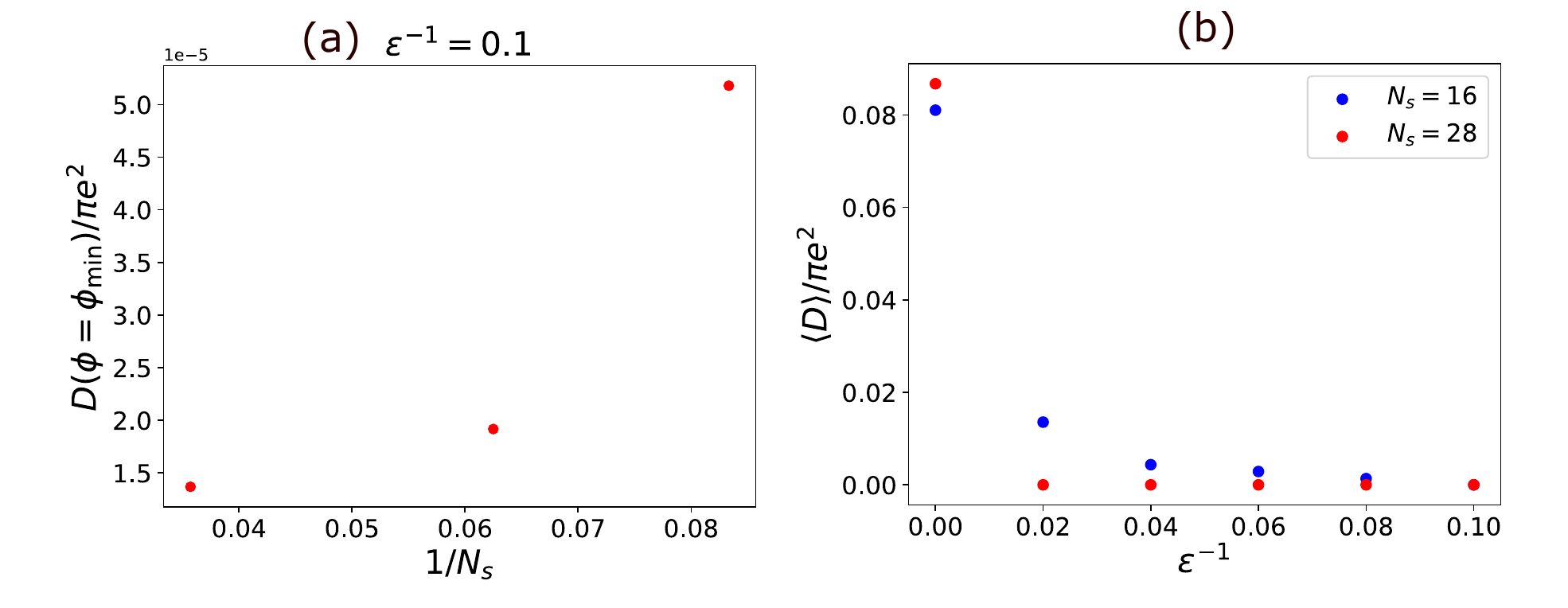}
    \caption{Drude weight for twisted $\rm{MoTe_2}$ \cite{reddy2023fractional} at twist angle $\theta = 2.0 ^\circ$. (a) The Drude weight calculated at the flux that minimizes the energy dispersion. (b) The average Drude weight ${\langle D \rangle}$ as a function of the inverse dielectric constant. $N_s$ denotes the number of sites. } 
    \label{fig:S_2}
\end{figure}
\textcolor{black}{\section{Landau levels modulated with a $C_4$ symmetric potential}
In this section, we study Landau levels modulated with a $C_4$ symmetric potential as an additional evidence for the vanishing Drude weight in the CFL phase. For the case here, we take the potential to enclose one flux quantum defining a unit cell that is spanned by the two vectors $\mathbf{R}_1 = \sqrt{2 \pi} l \> (1,0) $ and $\mathbf{R}_2 = \sqrt{2 \pi} l \> (0,1)$ giving rise to reciprocal lattice vectors, $\mathbf{g}_1 = g_0 (1,0)$ and $\mathbf{g}_2 = g_0 (0,1)$ with $g_0 = \frac{1}{\sqrt{2 \pi} l}$. We consider a modulating potential that is a sum of the lowest harmonics ,  $V(\mathbf{r}) = 2 V_0 [\cos(\frac{2 \pi}{a_0} r_x)  + \cos(\frac{2 \pi}{a_0} r_y)]$, giving rise to the following dispersion when projected to the lowest Landau level,
\begin{equation}
    \tilde{E}_0(\mathbf{k}) = -V_0 e^{-l^2 g_0^2/4} [\cos( l^2 g_0 k_x) + \cos( l^2 g_0 k_y)]
\end{equation} 
where $V_0$ denotes the strength of the potential. We study the interacting problem at half filling as a function of $V_0$ and the interaction strength $U_0$. In Fig. \ref{fig:square_supp}, we show the many-body spectrum in the CFL phase (Fig. \ref{fig:square_supp}(a)) and the FL phase (Fig. \ref{fig:square_supp})(b)). In the CFL phase, the momentum occupation of the many body ground state is expected to be uniform due to the absence of electronic Fermi surface. This is indeed the case as evident from Fig. \ref{fig:square_supp}(c). On the other hand, we find a clear Fermi surface in the momentum occupation in the FL phase (Fig. \ref{fig:square_supp}(d)). The finite-size scaling of the Drude weight in both phases is shown in Fig. (5) in the main text.}
\begin{figure}[t!]  
     \centering
     \textcolor{black}{
     \includegraphics[width = 0.8\linewidth]{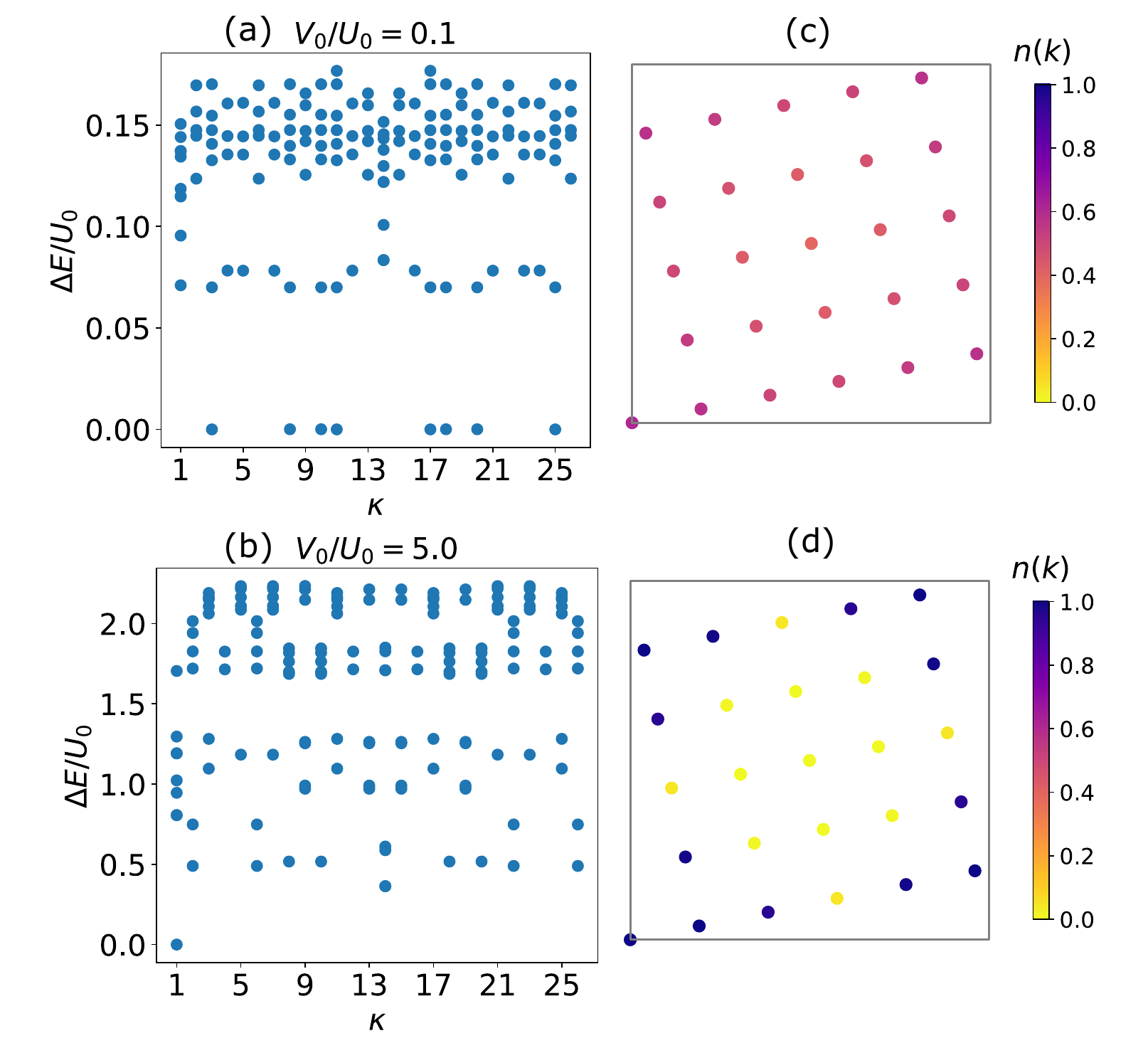}
     \caption{(a)-(b) The many-body spectrum in the CFL phase (a) and the FL phase (b) for modulated Landau levels with $C_4$ symmetric periodic potential \eqref{eq:dispersion_square} (c)-(d) Momentum space occupation number $n(\mathbf{k}) = \langle c^{\dagger}_{\mathbf{k}} c_{\mathbf{k}}\rangle $ of the many-body ground states in (a) the CFL phase and (b) the FL phase. $V_0/U_0$ denotes the ratio of the periodic potential strength to the interaction strength. Calculations are done at zero flux $\phi =0 $. $\kappa$ labels the total momentum sectors.} 
     \label{fig:square_supp}}
 \end{figure}

\begin{figure}[t!]  
    \centering
    \textcolor{black}{
    \includegraphics[width = 0.7 \linewidth]{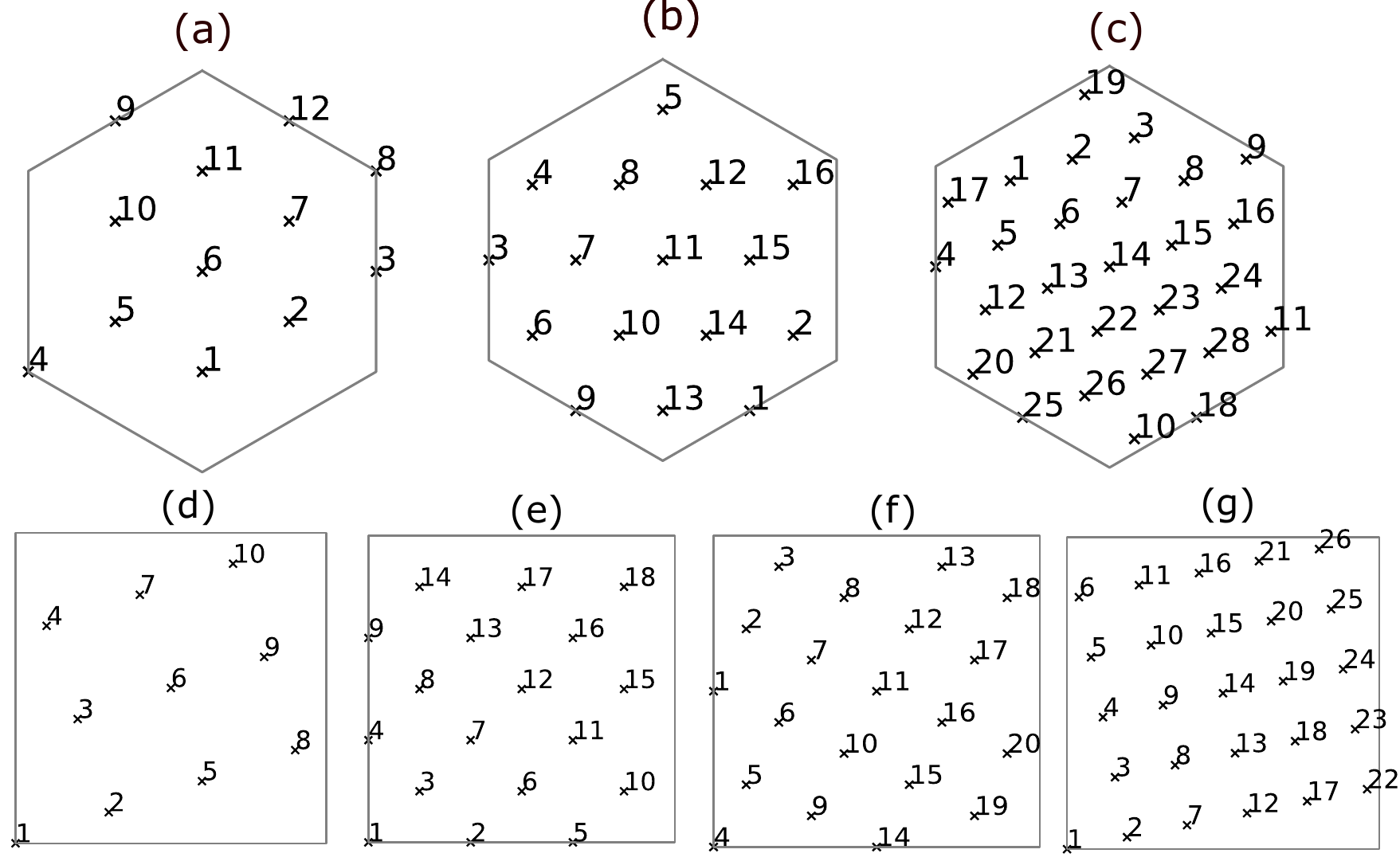}
    \caption{The momentum grids corresponding to the finite systems used in this work. Each momentum point is labeled by an index $\kappa$.} 
    \label{fig:S_3}}
\end{figure}

\section{Exact Diagonalization and Drude weight calculation}
In this paper, we use exact diagonalization (ED) on finite systems to study the interacting Hamiltonian. The finite systems are are spanned by two vectors,

\begin{equation}
\begin{aligned}
    \mathbf{L}_1 = m_1 \mathbf{a}_1 + n_1 \mathbf{a}_2 \\
\mathbf{L}_2 = m_2 \mathbf{a}_1 + n_2 \mathbf{a}_2
\end{aligned}
\end{equation} where $\mathbf{a}_1$ and $\mathbf{a}_2$ are two real space basis and $m_1,n_1,m_2,n_2$ are integers.

We work directly in momentum space. The momentum points live on a grid given by 
\begin{equation}
    \mathbf{k} = ( 2\pi \alpha + \phi_1) \> \mathbf{b}_1 + (2\pi \beta + \phi_2) \> \mathbf{b}_2
\end{equation}
where $\mathbf{b}_1$ and $\mathbf{b}_2$ are the basis vectors of the momentum grid, defined as $\mathbf{b}_i\cdot \mathbf{L}_j = \delta_{ij}$ and $\alpha,\beta$ are integers.  $(\phi_1,\phi_2) \in [0,2\pi)$ are the flux inserted through the cycles of the torus corresponding to generic twisted boundary conditions.

In Fig. \ref{fig:S_3}, we show the finite systems used in this work. We have only considered finite systems that are fully $C_6$ symmetric and have equal aspect ratio for a faithful representation of the models we studied. 

For a given system, the interacting Hamiltonian is diagonalized and the low-lying energy states are obtained as usual. We consider only threading a flux $\phi$ in one direction, say in the $\mathbf{L}_1$ cycle. Due to the equal aspect ratio of the clusters used, identical results are obtained when the flux is inserted in the other direction instead. 

In our work, we evaluate the Drude weight which is proportional to the second derivative of the ground state energy with respect to the flux, $\frac{\partial^2 E_0 (\phi)}{\partial \phi^2}$ where $E_0(\phi)$ is the ground state of the system obtained from the ED at flux $\phi$. To evaluate a single $\frac{\partial^2 E_0 (\phi)}{\partial \phi^2}$ , we use numerical finite-difference method $\frac{\partial^2 E_0 (\phi)}{\partial \phi^2} \approx (E_0(\phi + \delta\phi) - 2 E_0(\phi) + E_0(\phi - \delta \phi))/(\delta \phi^2) $ where $\delta \phi $ \textcolor{black}{is chosen to be between $\delta \phi = 2\pi/(20)$ and $ \delta \phi = 2\pi/(100)$.}

To evaluate the average Drude weight obtained by integrating $\frac{1}{2 \pi} \int d^2 \phi \frac{\partial^2 E_0 (\phi)}{\partial \phi^2}$, we notice first that if $E_0(\phi) $ is a smooth function of $\phi$ with no cusps, the integral vanishes identically since it's an integral of a smooth and periodic function over the periodicity interval. The only way the integral can have a finite value is when there exist cusps (discontinuities in the first derivative) due to level crossings. In such a case, we identify all the cusp points and the integral is calculated as
\textcolor{black}{
\begin{equation}
    \int d\phi \frac{\partial^2 E_0 (\phi)}{\partial \phi^2} =  \sum_{p}(\frac{\partial E_0(\phi)}{\partial \phi}|_{p_{+}} - \frac{\partial E_0(\phi)}{\partial \phi}|_{p_{-}})
\end{equation}}
where $p$ runs over all cusp points and $\pm$ refers to evaluating the first derivative from the left ($+$) or the right ($-$). By isolating all the cusp points, the energy $E_0(\phi)$ as a function of flux is a piecewise function defined over different intervals where in each interval $E_0(\phi)$ is smooth. We then fit $E_0(\phi)$ to a second order polynomial of $\phi$ and extract the first derivative $\partial E_0(\phi)/(\partial \phi)$ from the fitted polynomial. We find this fitting procedure to give more accurate results than evaluating the first derivative around the cusps using finite-difference.

\end{widetext}

\end{document}